\documentstyle[11pt,epsfig]{article}
\newcommand{\dd}{{\rm d}}
\newcommand{\url}[1]{[{\tt #1}]}
\begin{document}

\title{Variation of the constants in the late and early universe}

\author{Jean-Philippe Uzan}

\date{Laboratoire de Physique Th\'eorique, CNRS-UMR 8627, B\^at 210,\\
           Universit\'e Paris XI, 91415 Orsay cedex (France) \\
           and\\
           Institut d'Astrophysique de Paris, GReCO, 8bis bd Arago, 75014 Paris (France).\\
      10 september, 2004}
\maketitle

\begin{abstract}
Recent key observational results on the variation of fine
structure constant, the proton to electron mass ratio and the
gravitational constant are reviewed. The necessity to substantiate
the dark sector of cosmology and to test gravity on astrophysical
scales is also emphasized.
\end{abstract}
\maketitle
\section{Introduction}

The question of whether the constants of nature may be dynamical
goes back to Dirac~\cite{dirac37} who expressed, in his ``Large
Number hypothesis", the opinion that very large (or small)
dimensionless universal constants cannot be pure mathematical
numbers and must not occur in the basic laws of physics. He
stressed that the ratio between the gravitational and
electromagnetic forces between a proton and an electron, $Gm_{\rm
e}m_{\rm p}/e^2\sim10^{-40}$ is of the same order as the inverse
of the age of the universe in atomic units, $e^2H_0/m_{\rm e}c^3$.
He stated that these were not coincidences and that these big
numbers were not pure constants but reflected the state of our
universe. This led him to postulate that $G$ varies\footnote{Dirac
hypothesis can also be achieved by assuming that $e$ varies as
$t^{1/2}$. Indeed the choice depends the choice of units, either
atomic or Planck units. There is however a difference: assuming
that only $G$ varies violates the strong equivalence principle
while assuming a varying $e$ results in a theory violating the
Einstein equivalence principle. It does not mean we are detecting
the variation of a dimensionful constant but simply that either
$e^2/\hbar c$ or $Gm_{\rm e}^2/\hbar c$ is varying.} as the
inverse of the cosmic time.

Diracs' hypothesis is indeed not a theory and it was shown later
by Jordan~\cite{jordan} (see Ref.~\cite{u02} for details) that
varying constants can be included in a Lagrangian formulation as
new dynamical degree of freedom so that one gets both a dynamical
equation of evolution for this degree of freedom and a
modification with respect to the equations derived under the
hypothesis it is constant. Testing for their constancy is a
fundamental test of gravitation related to the local position
invariance. It was also realized~\cite{gr} that varying constants
may be associated to the existence of a composition dependent
fifth force and thus to a violation of the universality of free
fall.

In this review talk, I first emphasize the necessity to test
gravity on astrophysical scales. In particular, I will discuss
what can be learnt from the tests of the constancy of the
constants. While it is tested with increasing precisions in the
laboratory (see Ref.~\cite{submm} for a recent review), in the
Solar System and by the study of pulsar timing (see the
contribution by G. Esposito-Far\`ese in this volume), there exist
very few tests on astrophysical and cosmological
scales~\cite{u03}. Such tests are however required mainly because
the cosmic matter budget~\cite{budget} mostly relies on the
properties of gravitation, as well as the conclusion that about
96\% of the energy density of our universe is dark (including dark
matter and dark energy).

I will then review the recent observational and experimental
constraints on the variation of the fine structure constant, the
gravitational constant and the proton to electron mass ration. Let
us stress once again that we are only able to detect the variation
of dimensionless constants (see Refs.~\cite{eu03,okun91} for
discussions and I refer to Refs.~\cite{okun91,trialogue,ul} for
the distinction between fundamental parameters and fundamental
units, as well as for their role in the formulation of the laws of
physics).

To finish, I discuss some developments concerning the
phenomenology of varying constants both in the late and early
universe.

\section{Gravity on astrophysical scales}

\subsection{The dark sector and gravity}

Most cosmological observations now provide compelling evidences
that our universe is undergoing a late time acceleration phase,
the interpretation of which is still a matter of debate. In
particular, the Friedmann equations  for a universe filled  only
with pressureless matter (including dark matter) and radiation
\begin{equation}\label{eq1}
 H(z) = H_0 E(z)
 \quad\hbox{with}\quad
 E^2(z) = \Omega_{\rm m}^0(1+z)^3 + \Omega_{\rm r}^0(1+z)^4 +
 \Omega_K^0(1+z)^2
\end{equation}
cannot explain the current data~\cite{dark_revue}. Various ways to
face this fact have been considered but all lead to the
introduction of new degrees of freedom, referred to as {\it dark
energy}, in the cosmological scenario, either as new components of
gravitating matter or as new properties of gravity\footnote{Note a
third possibility. To infer the existence of this dark energy, we
interpret the cosmological observations in a given theoretical
frame which assumes e.g. symmetries for the spacetime. It may be
that the observations are not interpreted in the correct frame.}.

In the first approach, it is assumed that there exists new
gravitating components, beyond the standard model of particle
physics, while gravity is supposed to be accurately described by
general relativity. Many candidates such as a cosmological
constant, quintessence~\cite{qessence}, K-essence~\cite{kessence}
have been proposed (see e.g. Ref.~\cite{lambda} for a review).
From a cosmological point of view, these models are characterized
by their equation of state which can be reconstructed from the
function $E(z)$. Note that most late time observations (such as
diameter and luminosity distances or the growth factor of cosmic
structures) only depend on some combination of this function
$E(z)$.

The  other route is to allow for a modification of gravity which
means that {\it the long range force that cannot be screened is
assumed not to be described by general relativity}. Many such
models have been considered. For instance, a light scalar field
can couple to matter leading to scalar-tensor models of
quintessence~\cite{u99,stquint}. This scalar field responsible for
variation of the gravitational constant may also be, depending on
its couplings, at the origin of the variation of other constants
and of a violation of the universality of free fall (see
Ref.~\cite{u02,wetterich} for details). Other possibilities
include braneworld models. Higher dimensional models predict that
gravity should depart from its standard Newton behavior on {\it
small scales} and up to now this scale is constrained to be
smaller than $100-500\,\mu$m~\cite{submm}. Among braneworld
models, a class has also the feature to allow for deviations from
4-dimensional Einstein gravity on large scales. This is for
example the case of some multi-brane models~\cite{multi_brane},
multigravity~\cite{multigravity}, brane induced
gravity~\cite{induced_grav} or simulated gravity~\cite{simugrav}
where gravity is not mediated only by a massless graviton but
include a tower of massive gravitons.

\begin{figure}[ht]
  \centerline{\epsfig{figure=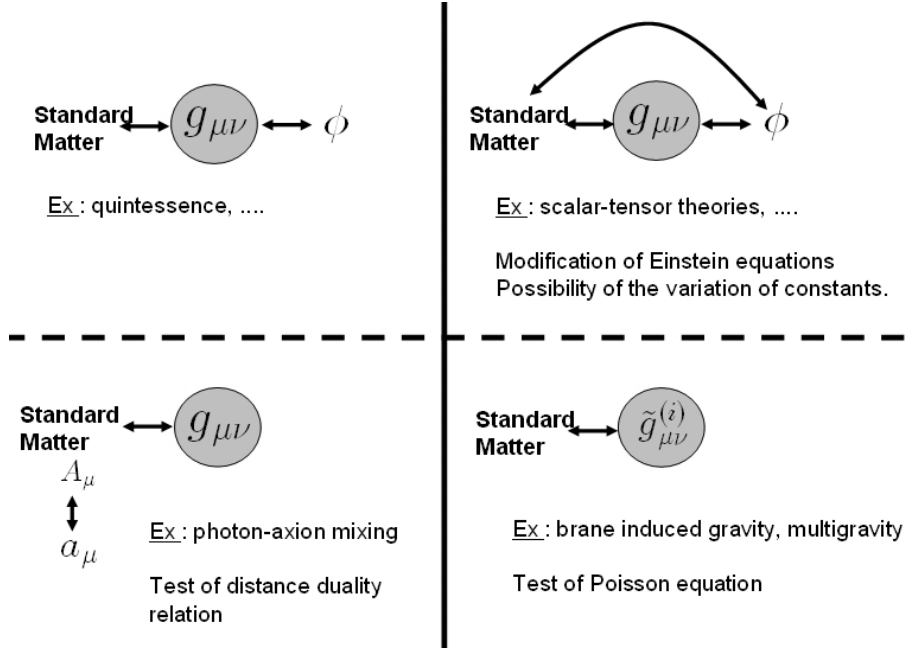,width=12cm}}
  \caption{Summary of the different classes of models and of the specific tests
 that can help distinguish between them (besides the equation of state
 and the growth of cosmic structures). The classes differ according
 to the kind of new fields and to the way they couple to the metric
 $g_{\mu\nu}$ and to the standard matter fields. Upper-left class consists of
 models in which a new kind of gravitating matter is introduced,
 e.g. quintessence, K-essence. In the upper-right
 class, a light field induces a long-range force so that gravity is
 not described by a spin-2 graviton only. This is the case of
 scalar-tensor theories of gravity. In this class, Einstein equations
 are modified and there may be a variation of the fundamental
 constants. Note also the possibility for this new field to couple
 differently to baryonic and dark matter. The lower-right class corresponds to models in which there
 may exist massive gravitons, such as in some class of braneworld
 scenarios. These models predict a modification of the Poisson
 equation on large scales. In the last class (lower-left), the
 distance duality relation may be violated. From Ref.~\cite{uam}.}
  \label{fig1}
\end{figure}

The dark sector plays an increasing role in cosmological models.
By testing the theory of gravity on astrophysical and cosmological
scales we will strengthen these conclusions. These tests will
contribute to substantiate the physics of the dark sector. Dark
matter may be more complicated than a pure collisionless gas and
dark energy may require to go beyond a pure scalar field
interacting only with gravity. Concerning the dark energy
phenomenology, the reconstruction of the function $E(z)$ will not
be sufficient to distinguish between many models\footnote{Indeed,
it is always possible~\cite{recon} to construct a scalar field
potential that will lead to the ``observed" $E(z)$ (on which most
of the observable -- diameter and angular distances, growth of
cosmic structures,...-- depend) so that the precise determination
of the late evolution history of our universe, even if very
constraining, will not allow us to determine the true nature of
dark energy.}. In Ref.~\cite{uam}, we proposed a classification of
these models and describes some of the specific signatures that
can discriminate between them. It is recalled in Fig.~\ref{fig1}.\\

From a theoretical point of view, string theory seems to be the
only known promising framework that can reconcile quantum
mechanics and gravity, even though it is not yet fully defined
beyond the perturbative level. One definitive prediction drawn for
the low-energy effective action is the existence of
extra-dimensions and of a scalar field, the dilaton, that couples
to matter~\cite{tv} and whose expectation value determines the
string coupling constant. It follows that the low-energy coupling
constants are in fact dynamical quantities. When the dilaton is
massless (or almost) it leads to 3 effects: (i) a scalar admixture
of a scalar component inducing deviations from general relativity
in gravitational effects, (ii) a variation of the couplings and
(iii) a violation of the weak equivalence principle.

From this perspective, testing gravity may also reveal the
existence of further gravitational fields or of extra-dimensions
and it opens an observational window on the low-energy limit of
string theory and/or on the stabilization of the dilaton and
extra-dimensions.

\subsection{What is tested and what should we test?}

Up to now, the observational status concerning the tests of
gravity is the following.
\begin{enumerate}
\item On Solar System size, the Newton law as well as the
universality of free fall are tested with a very good accuracy
(see Ref.~\cite{submm} for a summary of the constraints) .
\item On galactic scales, there are a
number of astrophysical constraints that a successful modification
of gravity will have to face (see e.g. Ref.~\cite{aguirre}). If
the modification of gravity has some relevance on galactic scales
then it will have to explain the flattening of the rotation curves
and to account for the dependence of the galaxy rotation curve on
the luminosity of the galaxy. This dependence is encapsulated in
the Tully-Fischer relation relating the luminosity of a spiral
galaxy to its asymptotic rotation velocity $L\propto
v_\infty^\alpha$, with $\alpha\sim4$. This sets severe constraints
on theories in which the cross-over scale with standard gravity is
fixed (see e.g. Ref~\cite{rotcurv1}) and favored theories where
this cross-over scale depends on the considered galaxy. Roughly,
one needs this crossover scale to behave as
\begin{equation}
 \ell_* \simeq \sqrt{\ell_0GM}/c
\end{equation}
where $\ell_0\sim10^{27}$m. On the other hand, the compatibility
between X-ray and strong lensing observations tends to show that
the Poisson equation holds up to roughly 2~Mpc~\cite{allen}.
\item On cosmological scales, there is at the moment no direct tests
of gravity. Indeed the growth of cosmological structure can put
some constraints but usually the observations entangle the
properties of the matter and gravity. Both the acceleration of the
universe and the variation of the constants (if confirmed) may be
indications of the break-down of general relativity in this
regime.
\end{enumerate}

\begin{figure}[ht]
  \centerline{\epsfig{figure=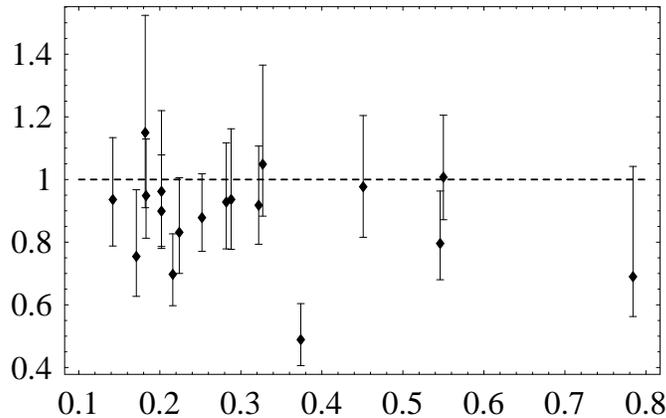,width=9cm}}
  \caption{$\eta={D_A}/{D_L}(1+z)^2$ as a function of the redshift for the 18 clusters
 of the Reese {\em et al.}~\cite{reese} catalog. The error bars
 include the observational error bars as determined by Reese {\em et
 al.}~\cite{reese} and the uncertainties in the cosmological parameters. Outlayers
 correspond to bimodal clusters that cannot be well fitted by a $\beta$-profile.
 This shows that there is no significant departure from the expected distance duality
 relation. From Ref.~\cite{uam}.}
  \label{fig2}
\end{figure}

{\it What should we test then?} General relativity is based on
Einstein equivalence principle that includes three hypothesis: (i)
local Lorentz invariance, (ii) local position invariance and (iii)
universality of free fall. If these hypothesis are valid, it is
thought that gravity is a geometric property of spacetime (see
e.g. Ref.~\cite{will}). We may thus aim to test for both the
Einstein equivalence principle and the field equations that
determine the geometric structure created by any mass
distribution. Some tests in these directions have been proposed in
the past years.
\begin{itemize}
\item {\it Lorentz invariance:} it can be tested through the propagation
of high energy particles. A summary of the constraints can be
found in e.g. Ref.~\cite{jacobson04}
\item {\it Poisson equation:} on sub-Hubble scales, the Einstein
equations in an expanding spacetime reduce to the Poisson equation
\begin{equation}
 \Delta\Phi = 4\pi G\rho a^2 \delta
\end{equation}
that relates the gravitational potential to the density contrast.
It was recently argued~\cite{ub01} that the comparison of galaxy
catalogs such as SDSS or 2dF and of weak lensing observations give
a direct test of this equation and can be extended up to 100~Mpc
(see also Ref.~\cite{white01}).
\item {\it Distance duality:} as long as photons travel on null
geodesics and the geodesic deviation equation holds, it can be
shown that there is a {\it reciprocity relation}, $r_S=r_O(1+z)$
between the source and observer area distances. Indeed $r_S$
cannot be measured so that this relation cannot be tested.
However, if the number of photons is conserved, it translates to a
{\it distance duality relation} between the luminosity and angular
distances, $D_L=D_A(1+z)^2$ that can be tested. This
proposition~\cite{bk03} has been tested~\cite{uam} using X-ray and
Sunayev-Zeld'ovich observations of galaxy clusters and no
departure from the standard expectation has been found (see
Fig.~\ref{fig2}). This can constrain e.g. models involving
photon-axion mixing~\cite{photgam}.
\item {\it Growth of cosmic structures:} the growth factor of cosmic
structures is sensitive to the equation of state of the matter
that drives the expansion of the universe~\cite{bb01}. It was
recently proposed to use the skewness and higher moments of the
density field induced during the non-linear Newtonian clustering
to test for the existence of a long-range Yukawa
force~\cite{sealfon04}. Similarly, it was argued that this
skewness was sensitive to the coupling of dark energy to dark
matter~\cite{amendola04,reis04}.
\item {\it Testing for the constancy of the (non-gravitational) constants} is a direct test
of the local invariance position\footnote{The local position
invariance implies that the (non-gravitational) laws of physics
determined locally take the same form at any spacetime point. This
demands that some  uniformity is necessary for reasonable
predictions to be made about distant part of the universe. This
can also be called a {\it local predictability
assumption}~\cite{gfr1}. Indeed, it does not exclude theories in
which the gravitational constant varies. Assuming we can determine
its law of variation, it just implies that local physics is  more
complex than we might have thought initially. Again, we see the
interplay between cosmological tests and local physics.} that
makes it a test of Einstein equivalence principle and of the
strong equivalence principle when the gravitational constant is
considered. It has been investigated by many methods and is the
subject of the following section. Note that if violated, one
expects also a violation of the universality of free fall mainly
because the self-energy is composition dependent and gravitates.
\end{itemize}

\begin{figure}[ht]
  \centerline{\epsfig{figure=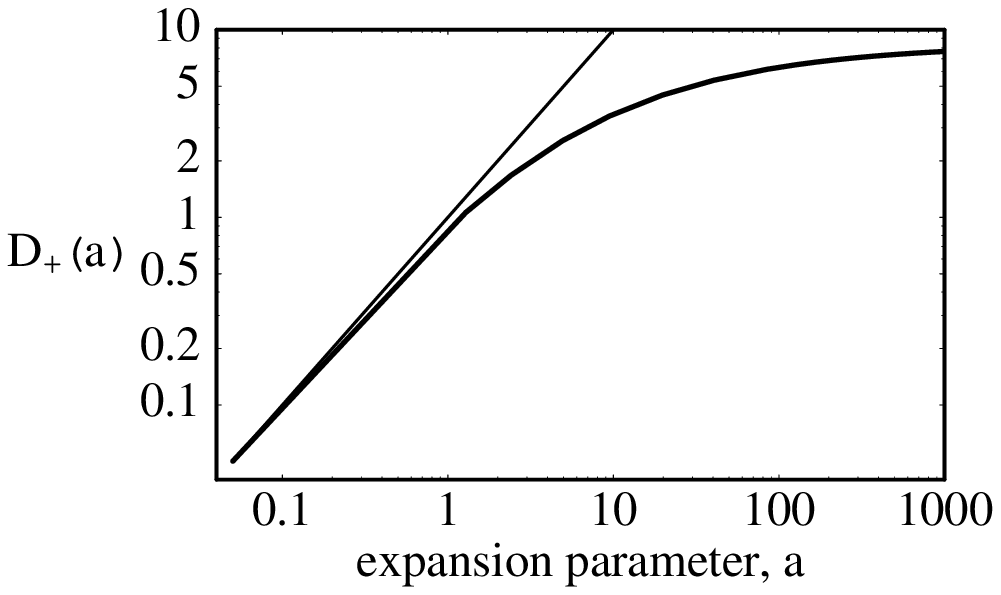,width=8cm}
   \epsfig{figure=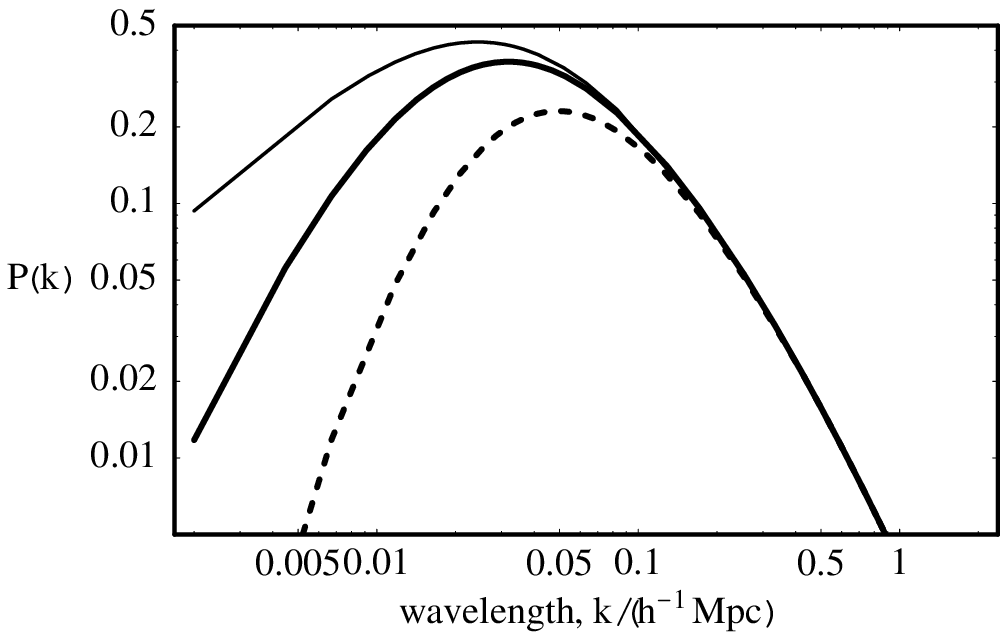,width=8cm}}
  \caption{In a theory in which gravity switches from a standard
 four dimensional gravity to a five dimensional gravity above a
 crossover scale of $r_s=50h^{-1}$~Mpc, there are different cosmological
 implications concerning the growth of cosmological perturbations. Since
 gravity becomes weaker on large scales, fluctuations stop growing
 [left panel]. It implies [right panel] that the density contrast power spectrum (thick
 line) differs from the standard one (thin line) but, more important, from
 the gravitational potential power spectrum (dash line). From Ref.~\cite{ub01}.}
  \label{fig3}
\end{figure}

\section{Update on the constraints on the variation of the constants}

Since the claim~\cite{webb99} that the fine structure constant,
$\alpha$, may have been smaller in the past, there have been a
tremendous increase in the interest of the fundamental constants
of nature and to whether there are really constant. Various
reviews~\cite{u02,damour03} exist and details and more references
can be found in Ref.~\cite{u02}

We review the recent developments concerning mainly the
constraints on the variation of the fine structure constant, as
well as the proton to electron mass ration, $\mu$, and the
gravitational constant, $G$.

\subsection{Laboratory constraints}

There have been some marked improvements on the constraints on the
variation of $\alpha$ in laboratory experiments.  These methods
are based on the comparison of atomic clocks using different types
of transitions in different atoms. According to the comparison,
one can constrain the variation of some combination of fundamental
constants. For instance, the hyperfine transition frequency of
alkali can be approximated by~\cite{karshenboim03}
\begin{equation}
 \nu\propto \alpha^2\frac{\mu}{\mu_N} \frac{m_{\rm e}}{m_{\rm p}}R_\infty c F_{\rm
 rel}(Z\alpha)
\end{equation}
where $\mu$ is the magnetic moment of the nucleus, $\mu_N$ the
nuclear magneton, $R_\infty$ the Rydberg constant and $F_{\rm
rel}$ a relativistic function~\cite{karshenboim03} which strongly
increases with the atom number $Z$ (e.g. $\dd\ln F_{\rm
rel}(Z\alpha)/\dd\ln\alpha \simeq 0.74$ for $^{133}$Cs and 0.30
for $^{87}$Rb).

The comparison of hyperfine transitions in $^{87}$Rb and
$^{133}$Cs over a period of about 4 years took advantage of this
sharp variation to show~\cite{marion02} that $\dd\ln(\nu_{\rm
Rb}/\nu_{\rm Cs})/\dd t = (0.2 \pm 7.0 )\times 10^{-16}\,{\rm
yr}^{-1}$ at 1$\sigma$. Neglecting possible changes in the
amplitude of the weak and strong interactions and thus in the
nuclear magnetic moments, it translates to
\begin{eqnarray}
 \dot\alpha/\alpha = (-0.4\pm16)\times10^{-16}\,{\rm yr}^{-1}.
\end{eqnarray}

Another experiment~\cite{bize03} comparing an electric quadrupole
transition in $^{199}$Hg$^+$ to the ground-state hyperfine
splitting of $^{133}$Cs over a 3 years period showed that
$|\dd\ln(\nu_{\rm Hg}/\nu_{\rm Cs})/\dd t| <7.0\times
10^{-15}\,{\rm yr}^{-1}$. This constrains the time variation of
$g_{\rm Cs}(m_{\rm e}/m_{\rm p})\alpha^{6.0}$ so that
\begin{eqnarray}
 \left|\dot\alpha/\alpha\right| < 1.2\times10^{-15}\,{\rm yr}^{-1}
\end{eqnarray}
if both the gyromagnetic factor $g_{\rm Cs}$ and $m_{\rm e}/m_{\rm
p}$ are assumed constant.

The comparison~\cite{fischer03} of the absolute $1S-2S$ transition
in atomic hydrogen to the ground state of cesium combined with the
results of Refs.~\cite{marion02,bize03} yields the two independent
constraints
\begin{eqnarray}
 \dot\alpha/\alpha = (-0.9\pm2.9)\times10^{-16}\,{\rm
 yr}^{-1},\qquad
 \dd\ln(\nu_{\rm Rb}/\nu_{\rm Cs})/\dd t = (-0.5\pm1.7)\times
10^{-15}\,{\rm yr}^{-1}.
\end{eqnarray}

The comparison~\cite{peik03} of optical transitions in
$^{171}$Yb$^+$ to a cesium atomic clock at two times separated by
2.8 years has shown that $\dd\ln(\nu_{\rm Yb}/\nu_{\rm Cs})/\dd t
= (-1.2\pm 4.4) \times 10^{-15}\,{\rm yr}^{-1}$ which translates
to
\begin{eqnarray}
 \dot\alpha/\alpha = (-0.3\pm2.0)\times10^{-15}\,{\rm yr}^{-1}.
\end{eqnarray}
These methods allow to set very sharp local constraints and, as
illustrated by the results of Ref.~\cite{fischer03}, they can be
combined to set independent constraints on various constants.

\subsection{Geochemical constraints}

Sharp constraints on the time variation of the fine structure
constant can be also obtained from the Oklo phenomenon and from
the study of the lifetimes of long-lived nuclei.\\

The Oklo phenomenon is a natural nuclear reactor that operated
during 200,000 years approximatively two billion years ago, that
is at a redshift $z\sim0.14$. The isotopic abundances of the
yields give access to informations about the nuclear rates at that
time. One of the key quantity measured is the ratio
${}^{149}_{62}{\rm Sm}/{}^{147}_{62}{\rm Sm}$ of two light
isotopes of samarium which are not fission products. This ratio of
order of 0.9 in normal samarium, is about 0.02 in Oklo ores. This
low value is interpreted by the depletion of ${}^{149}_{62}{\rm
Sm}$ by thermal neutrons to which it was exposed while the reactor
was active. The capture cross section of thermal neutron by
$^{149}{\rm Sm}$
\begin{equation}
 ^{149}{\rm Sm} + n \rightarrow ^{150}{\rm Sm} + \gamma
\end{equation}
has a resonant energy $E_r\simeq0.0973$~eV, which is a consequence
of a near cancellation between electromagnetic and strong
interactions~\cite{oklo1}. A detailed analysis and model of the
samarium nuclei lead~\cite{oklo2}, assuming that the variation of
$E_r$ is due only to the $\alpha$ dependence of the
electromagnetic energy, to the constraint
\begin{equation}
 \Delta\alpha/\alpha = (0.15\pm1.05)\times10^{-7}
\end{equation}
at 2$\sigma$. In particular, the accuracy of the method can be
understood by comparing the resonant energy $E_r\sim0.1$~eV to its
sensitivity to a variation of $\alpha$, $\dd
E_r/\dd\ln\alpha\sim-1$~Mev so that variation smaller than
$\Delta\alpha/\alpha\sim0.1\,{\rm eV}/1\,{\rm MeV}\sim 10^{-7}$
are expected.

It was later pointed out~\cite{oklo3} that there may be two ranges
of solutions compatible with the Oklo data
\begin{equation}
 \Delta\alpha/\alpha = (-0.8\pm1.0)\times10^{-8},\qquad
 \Delta\alpha/\alpha = (8.8\pm0.7)\times10^{-8},
\end{equation}
the second branches being disfavored by the analysis of the
isotopic ratio of gadolinium.

Recently, the assumption that the low energy neutron spectrum is
well described by a Maxwell-Boltzmann distribution was
investigated~\cite{oklo4}. The effect of a variation of the
strange quark mass was studied~\cite{oklo5} to show that $m_{\rm
s}/\Lambda_{\rm QCD}$ has varied by less than $1.2\times10^{-10}$
while assuming all fundamental couplings to vary independently
led~\cite{oklo6} to the more stringent limit
$\Delta\alpha/\alpha<(1-5)\times10^{-10}$.\\

Radioactive decay lifetimes can also be used once the
$\alpha$-dependence of the decay rate is known. For instance, the
lifetime of a $\beta$- decay nuclei scales as
\begin{equation}
 \lambda\sim\Lambda(\Delta E)^p G_F^2\alpha^s
\end{equation}
where $\Delta E$ is the decay energy and $s$ the sensitivity. Many
nuclei were used but the sharpest constraint was obtained from the
$\beta$-decay of rhenium to osmium by electron emission
\begin{equation}
{}^{187}_{75}{\rm Re}\longrightarrow{}^{187}_{76}{\rm
Os}+\bar\nu_e+e^-,
\end{equation}
first considered by Peebles and Dicke~\cite{os1}. Interestingly,
due its low decay energy --about 2.5 keV-- the sensitivity of
rhenium is $s\sim-18,000$ so that a variation of $\alpha$ of order
$10^{-2}\%$ induces a variation decay energy of order of the keV.

The analysis of new meteorite and laboratory led to~\cite{os2}
\begin{equation}
\Delta\alpha/\alpha = (8\pm16)\times10^{-7}
\end{equation}
over the last 4.5 Gyr, which corresponds to $z\simeq0.45$.

There is a caveat to this method that is not so direct: the ratio
Re/Os is measured in iron meteorite the age of which is not
determined directly. Models of formation of the solar system tend
to show that iron meteorites and angrite meteorites form within
the same 5 million years. The age of the latter can be estimated
from the $^{207}$Pb-$^{206}$Pb method which gives 4.558 billion
years. Besides, this constraint holds on the averaged value of
$\alpha$ over the 4.5 past billion years.

\subsection{Cosmological constraints}

On cosmological scales, the observation of the cosmic microwave
background (CMB) anisotropies and of the abundances of the light
elements produced during the big-bang nucleosynthesis (BBN) allow
to set constraints of order $10^{-2}$ on the variation of
$\alpha$.\\

Changing the fine structure constant modifies the strength of the
electromagnetic interaction and thus its only effect on CMB
anisotropies arises from the change in the differential optical
depth of photons due to the Thomson scattering, $\dot\tau=x_{\rm
e}n_{\rm e}c\sigma_{\rm T}$, which enters in the collision term of
the Boltzmann equation describing the evolution of the photon
distribution function and where $x_{\rm e}$ is the ionization
fraction (i.e. the number density of free electrons with respect
to their total number density $n_{\rm e}$).  The first dependence
of the optical depth on the fine structure constant arises from
the Thomson scattering cross-section given by $\sigma_{\rm
T}=({8\pi}/{3})({\hbar^2}/{m_{\rm e}^2c^2})\alpha^2$. The second,
and more subtle dependence, comes from the ionization fraction. A
variation of the fine structure constant can thus be thought of as
considering a delayed recombination model.

Early works~\cite{avelino00,battye01,avelino01} based on BOOMERanG
and MAXIMA data tend to show that the fit to CMB data are improved
by allowing $\Delta\alpha\not=0$ while Landau {\em et
al.}~\cite{landau01} concluded from these data imply, assuming
spatially flat models with adiabatic primordial fluctuations, that
$-0.14<\Delta\alpha/\alpha<0.03$ at $2\sigma$ level. The recent
analysis~\cite{cmb1} of the WMAP data (see the contribution by G.
Rocha for details) gave the
\begin{equation}
 \Delta\alpha/\alpha=(-1.5\pm3.5)\times10^{-2}
\end{equation}
at $z\sim10^3$. Note that he variation of the gravitational
constant can also have similar effects on the CMB~\cite{ru02}. In
conclusion, constraints of order $1\%$ on the variation of
$\alpha$ can be obtained from the CMB only if the
cosmological parameters are independently known.\\

BBN theory predicts the production of the light elements in the
early universe the abundances of which rely on a fine balance
between the expansion of the universe and the weak interaction
rates which controls the neutron to proton ratio at the onset of
BBN. Basically, the abundance of helium-4 is given by
\begin{equation}
 Y_p=2\frac{(n/p)_{\rm f}\exp(-t_{\rm N}/\tau)}{1+(n/p)_{\rm f}\exp(-t_{\rm N}/\tau)}
\end{equation}
where $(n/p)_{\rm f}=\exp{(-Q/kT_f)}$ is the neutron to proton
ratio at the freeze-out time determined by $G_F^2(kT_{\rm
f})^5=\sqrt{GN}(kT_{\rm f})^2$, $N$ being the number of
relativistic degrees of freedom; $Q=m_{\rm n}-m_{\rm p}$, $\tau$
is the neutron lifetime, $G_F$ the Fermi constant and $t_{\rm N}$
the time after which the photon density becomes low enough for the
photo-dissociation to be negligible. As a conclusion, the
predictions of BBN involve a large number of fundamental
constants.

A change in $\alpha$ affects directly $Q$ and was
modelled~\cite{bbn1,bbn2} as $Q/\Lambda_{\rm QCD}= a\alpha+ b
v/\Lambda_{\rm QCD}$ where $v$ determines the weak scale and $a$
and $b$ are two numbers. Roughly, this implies that $\Delta
Y/Y\sim-\Delta Q/Q\sim0.6\Delta\alpha/\alpha$. On this basis, one
can set the constraint $|\Delta\alpha/\alpha| <
5\times10^{-2}$~\cite{bbn1,bbn2}, confirmed by a recent
analysis~\cite{bbng2} which gives
$|\Delta\alpha/\alpha|<6\times10^{-2}$. This does not take the
effect of the fine structure constant in the Coulomb barriers and
in the cross-sections, which was investigated in Ref.~\cite{bbn3}.
Recently, the effect of seven parameters
$(G,\alpha,v,m_e,\tau,Q,B_d)$, later related to the six constants
$(G,\alpha,v,m_e,m_u,m_d)$ was taken into account and led to the
constraint~\cite{bbn4}
\begin{equation}
 \Delta\alpha/\alpha=(6\pm4)\times10^{-4}.
\end{equation}
The effect of the strange quark mass was also
investigated~\cite{oklo5} and it was claimed that $m_{\rm
s}/\Lambda_{\rm QCD}$ has varied by less than $6\times10^{-3}$
since BBN.

\subsection{Astrophysical constraints}

Most of the excitement over the possibility of the fine structure
constant variation arises from the observation of distant quasar
absorption systems. The method is based on the comparison of
absorption spectra to laboratory spectra.

Initially, the method was based on alkali doublets, the splitting
of which gives access to the fine structure constant,
$\Delta\nu/\bar\nu\propto\alpha^2$. The analysis of Si~IV on 21
systems gave~\cite{murphy03b} $\Delta\alpha/ \alpha =
(-0.5\pm1.3)\times10^{-5}$ for $2\leq z\leq 3$. The most recent
constraint~\cite{pp} has been obtained from the analysis of SiIV
in 15 systems of the VLT/UVES sample, improving the former
constraint by a factor 3,
\begin{equation}
 \Delta\alpha/\alpha=(0.15\pm0.43)\times10^{-5},\qquad 1.59\leq z\leq 2.92.
\end{equation}
It is to be noted that none of the analysis based on the alkali
doublet method exhibit a hint of variation of $\alpha$.

\begin{figure}[h]
  \centerline{\epsfig{figure=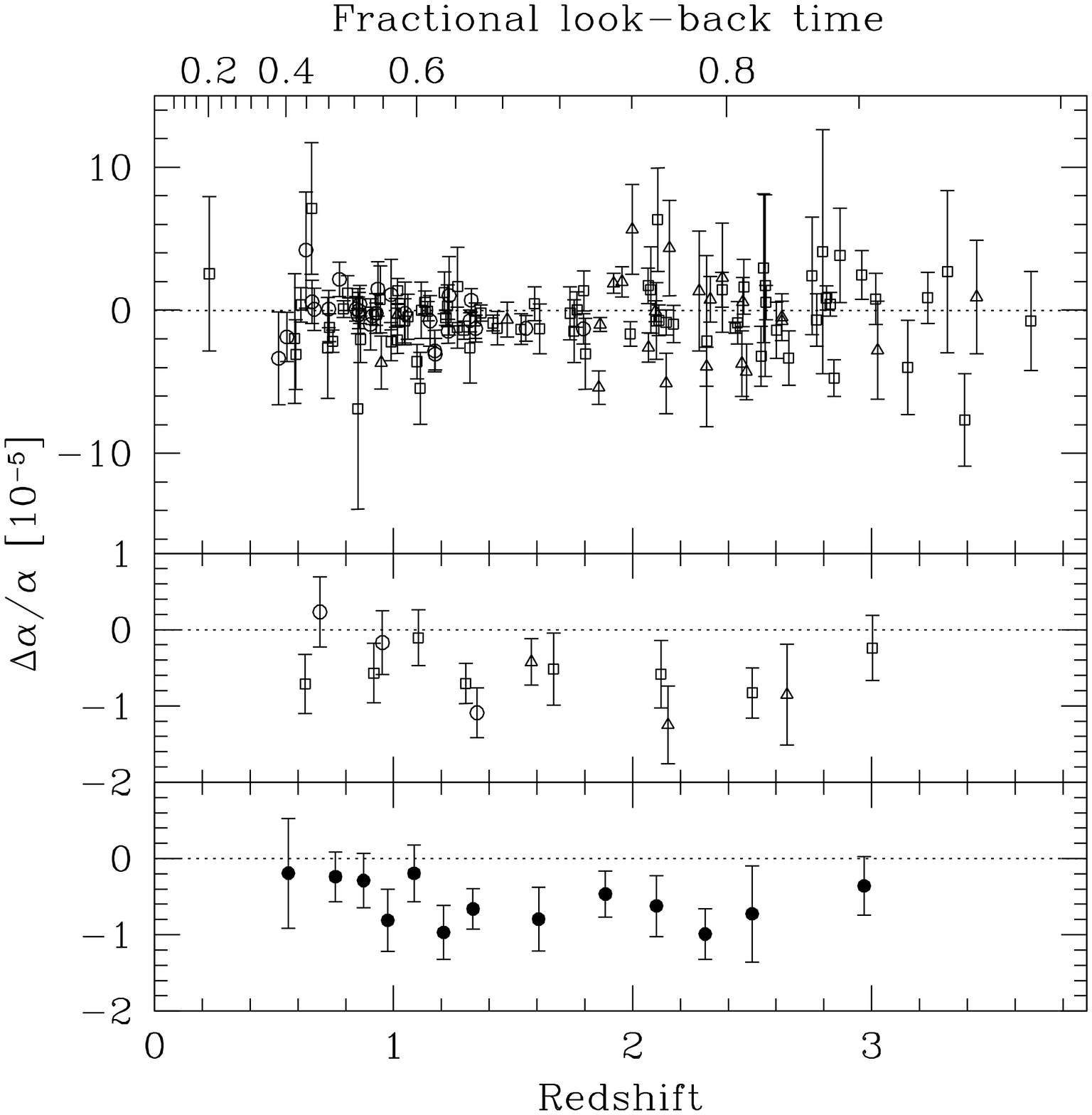,width=8cm}
   \epsfig{figure=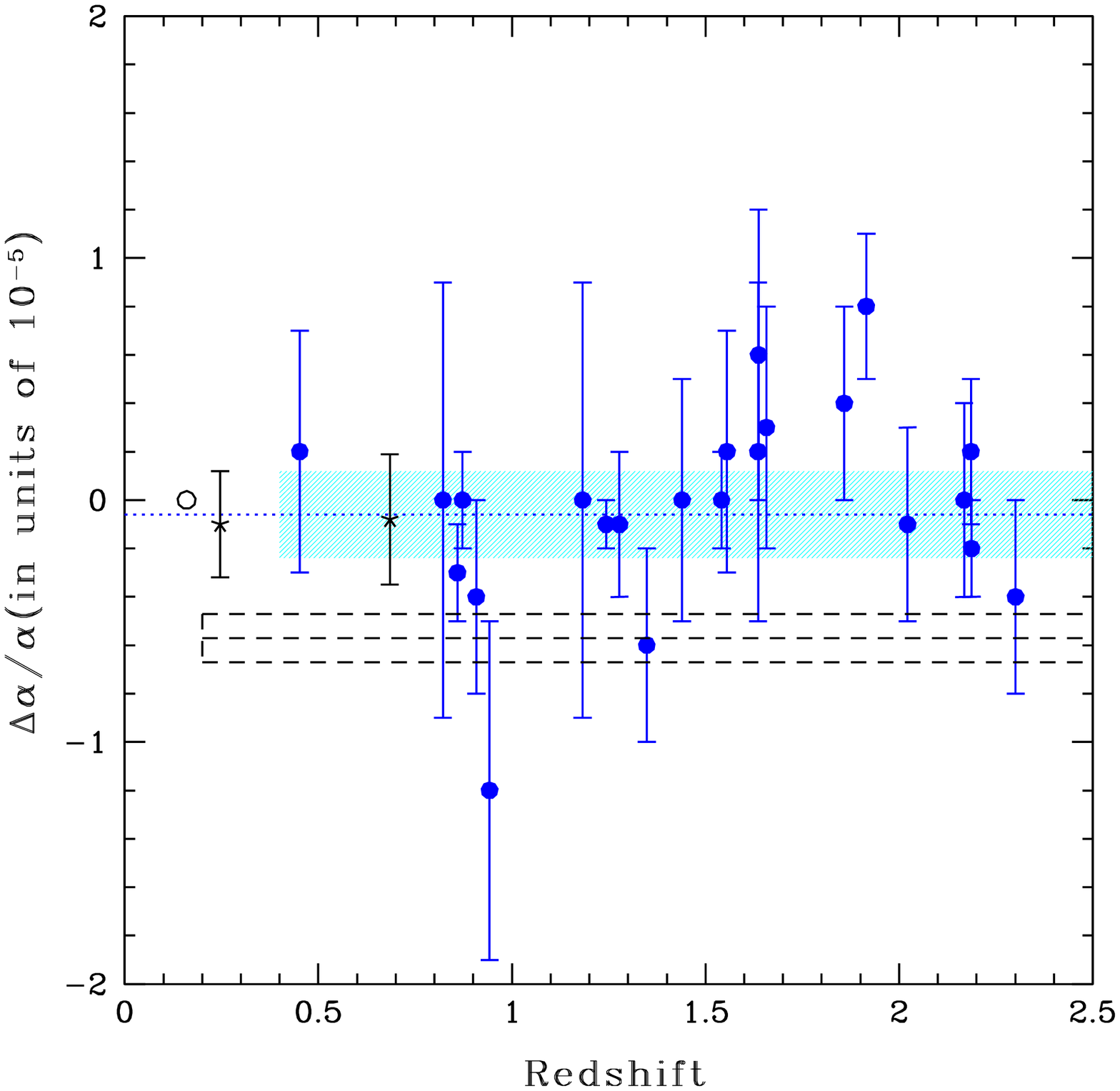,width=8cm}}
  \caption{The
  analysis of quasar spectra on the variation of the fine structure
  constant from (left) the Keck/Hires data~\cite{murphy03} and (right)
  the VLT/UVES data~\cite{chand04}.  The left plot depicts their
  previous low redshift (open circles), previous high redshift (open
  triangles) and new (open squares) samples. The raw results with
  $1\sigma$ error bar are shown on the top panel while the middle
  panel shows the results with an arbitrary bining and the bottom
  panel combines the three samples.  The right plot present the
  results from the VLT/UVES data~\cite{chand04}.  The dash lines
  represent the weighted mean and $1\sigma$ range from the analysis of
  Ref.~\cite{murphy03} while the dash region marks the weighted mean
  and its $3\sigma$ error bars.}  \label{fig4}
\end{figure}

The many multiplet (MM) method proposed by Webb {\it et
al.}~\cite{webb99,webb01} was aimed at increasing the precision of
the AD method by correlating  several transition lines from
various species in order to reach a sensitivity of order
$10^{-6}$. In particular, one can compare line shifts of element
which are sensitive to variation in $\alpha$ with those that are
not. At low redshift ($z\leq1.8$), the results lie mainly on the
comparison of Fe to Mg while at higher redshift they lie on the
comparison of Fe and Si. The latest analysis~\cite{murphy03} of
the Keck/Hires data (see Fig.~\ref{fig4}) based on 128 systems in
the range $0.5\leq z\leq 3$ points toward a lower value of
$\alpha$ in the past
\begin{equation}
 \Delta\alpha/\alpha=(-0.54\pm0.12)\times10^{-5},\qquad 0.5\leq z\leq 3
\end{equation}
as previous analysis did~\cite{webb99,webb01}. A detailed budget
of the errors was done and a possible systematic effects was
looked at but none was exhibited. Note also that new synthetic
atomic spectra were produced~\cite{spectra}.

Recent observations from the VLT/UVES using the same MM method
have not been able to duplicate this
result~\cite{chand04,srianand04} (see Fig.~\ref{fig4}). The
analysis by Chand {\em et al.}~\cite{chand04} is mainly based on
the analysis of Fe and Mg in 23 systems toward 18 QSOs in the
range $0.4\leq z\leq2.3$ because they apply some selection
criteria to the lines. In particular (1) they kept only species
with similar ionization potentials (MgII, FeII, SiII and AlII) so
that they are most likely to originate from similar regions in the
cloud, (2) absorption lines that are contaminated by atmospheric
lines were rejected, (3) they put a threshold on the column
density so that all FeII multiplets are detected at 5$\sigma$, (4)
they checked that the anchor are not saturated (Mg~I and II are
fairly insensitive to variation of $\alpha$) and (5) they excluded
strongly saturated systems with large velocity spread. They
concluded~\cite{chand04,srianand04} that
\begin{equation}
 \Delta\alpha/\alpha=(-0.6\pm0.6)\times10^{-6},\qquad 0.4\leq z\leq 2.3
\end{equation}
at 1$\sigma$. The analysis of a single quasar~\cite{quast03} also
gave
\begin{equation}
 \Delta\alpha/\alpha=(-0.1\pm1.7)\times10^{-6},\qquad z=1.15
\end{equation}
mainly from the analysis of Fe lines. These results were further
confirmed by the analysis of Fe~II lines in an absorption system
at $z=1.839$ toward the quasar Q1101-264~\cite{lev04}
\begin{equation}
 \Delta\alpha/\alpha=(2.4\pm3.8)\times10^{-6},\qquad z=1.839.
\end{equation}
The combined Fe~II sample also gives
$\Delta\alpha/\alpha=(0.4\pm1.5)\times10^{-6}$ for two systems at
$z=1.15$ and $1.839$.

Both results~\cite{murphy03,chand04} are sensitive to the isotopic
abundances of magnesium and assume a solar ratio
$^{24}$Mg:$^{25}$Mg:$^{26}$Mg=79:10:11. It is however commonly
assumed that heavy magnesium isotopes are absent in low
metallicity environments such as absorption clouds. Assuming a
ratio 1:0:0 would have led to a more significant detection of
$\Delta\alpha/\alpha = (-0.98\pm0.13) \times10^{-5}$ and
$\Delta\alpha/\alpha = (-0.36\pm0.06) \times10^{-5}$ respectively
for the Keck/Hires~\cite{murphy03} and VLT/UVES~\cite{chand04}
data. This has led to a new interpretation~\cite{ashen04} of the
MM results in which the variation of $\alpha$ is explained by an
early nucleosynthesis of $^{25}$Mg and $^{26}$Mg. Interestingly a
ratio~\cite{chand04} ($^{25}$Mg+$^{26}$Mg)/$^{24}$Mg =
$0.62\pm0.05$ and $0.30\pm0.01$ can explain respectively the
Keck/Hires~\cite{murphy03} and VLT/UVES~\cite{chand04} data. A
model of nucleosynthesis in which $^{25}$Mg and $^{26}$Mg are
produced by intermediate mass ($4-6M_\odot$) in their asymptotic
giant branch was proposed. They can reach a temperature larger
than $7\times10^7$K so that proton capture processes in the Mg-Al
cycle are effective enough. This hypothesis can be tested by
looking at other heavy elements produced by these intermediate
mass stars.

To finish, let us mention the use of O~III emission lines of
quasars and galaxies. The analysis~\cite{bahcall} of 42 quasars
from SDSS early data release and of 165 quasars of SDSS data
release 1 led respectively to the two constraints
\begin{equation}
 \Delta\alpha/\alpha=(0.7\pm1.4)\times10^{-4},\qquad
 \Delta\alpha/\alpha=(1.2\pm0.7)\times10^{-4}
\end{equation}
in the range $0.16\leq z\leq0.8$. The use of the OH microwave
transition was also proposed~\cite{darling1} and the preliminary
analysis of the quasar PKS1412+135 led to~\cite{darling2}
\begin{equation}
 \Delta\alpha/\alpha=(0.51\pm1.26)\times10^{-4},\qquad
 z=0.2467.
\end{equation}
These methods can be extended to higher redshifts ($z\sim3-5$) and
have different systematics compared to the MM and AD methods which
makes them complementary. Also, the analysis of OH, combined with
HCO$^+$ lines gave the simultaneous bounds~\cite{kc}
$\Delta\alpha/\alpha = (-0.38\pm2.2)\times10^{-3}$, $\Delta\mu/\mu
=(-0.27\pm1.6)\times10^{-3}$ and $\Delta g_p /g_p
=(-0.77\pm4.2)\times10^{-3}$ at z=0.68.

\subsection{Other constants}

\subsubsection{Proton to electron mass ratio}

The observation of vibro-rotational transitions of H$_2$ in damped
Lyman-$\alpha$ systems allows to constrain the variation of the
proton to electron mass ratio, $\mu=m_{\rm p}/m_{\rm e}$. The
spectral lines of the quasar are usually translated into a reduced
redshift, defined as $\zeta_i\equiv (z_i - \bar z)/(1+\bar z)$
where $\bar z$ is the averaged redshift of the absorption system.
It can be shown that $\zeta_i = ({\Delta\mu}/{\mu})K_i$ where the
$K_i$ are sensitivity coefficients.

Various constraints have been obtained since 1975 all showing no
hint of variation. In particular, the analysis of 83 absorption
lines~\cite{mu0} gave the limit
\begin{equation}
 \Delta\mu/\mu = (-7.5\pm9.5)\times10^{-5}
\end{equation}
at a $2\sigma$ level. More recently, the analysis of
vibro-rotational lines of molecular hydrogen for the two quasars
Q1232+082 ($z=2.3377$) and Q0347-382 ($z=3.0249$) tend to
show~\cite{mu1} that
\begin{equation}
 \Delta\mu/\mu = (-5.7\pm3.8)\times10^{-5}
\end{equation}
at $1.5\sigma$. It has recently been revised~\cite{mu4} to
\begin{equation}
\Delta\mu/\mu = (2.97\pm0.74)\times10^{-5}.
\end{equation}
It was shown that the limiting factor of the analysis was the
precision of the determination of the spectra in the laboratory.
New data for the transition wavelengths of H$_2$ Lyman and Werner
bands have been obtained with an accuracy of
$5\times10^{-8}$~\cite{mu2}. The reanalysis of the published
spectra then led to $\Delta\mu/\mu=(-0.5\pm3.0)\times10^{-5}$ at
$2\sigma$ level, confirming, once again, that the determination of
the laboratory spectra is the key of the debate on a possible
variation of $\mu$.

\subsubsection{Gravitational constant}

Few new works concern the gravitational constant. Let us note a
new analysis of the BBN~\cite{bbnG} that tends to show that
\begin{equation}
 \Delta G/G = 0.01^{+0.20}_{-0.16}
\end{equation}
at 68\% C.L. It was also shown~\cite{bbng2} that the variation of
$G$ is correlated to the extra-number of relativistic degrees of
freedom through $\Delta G/G =7\delta N/43$. It follows that the
combined analysis of new $^4$He and WMAP data implies
\begin{equation}
 -0.10 < \Delta G/G < 0.13
\end{equation}
A recent analysis of the secular variation of the period of
nonradial pulsations of the white dwarf G117-B15A shows~\cite{gwd}
that $0<\dot G/G < 4.0\times10^{-11}\,{\rm yr}^{-1}$ at 2$\sigma$,
which is of the same order of magnitude of previous independent
bounds (see also Ref.~\cite{gwd2}).

More important, in the Solar System, the measurement of the
frequency shift of radio photon to and from the Cassini spacecraft
improved the constraint on the post-Newtonian parameter $\gamma$
to~\cite{ssG} $\gamma-1 = (2.1\pm2.3)\times10^{-5}$. This can be
translated, in a model dependent way, into a constraint on the
time variation of $G$. For instance in a Brans-Dicke theory in a
matter dominated universe, it implies $|\dot
G/G|\leq10^{-14}\,\mathrm{yr}^{-1}$.

\section{Theoretical motivations and modelling}

\subsection{From string theory to phenomenology}

Most higher dimensional theories, such as Kaluza-Klein and string
theories, imply that dimensionless constants are
dynamical~\cite{tv,polchinski}. For example, in type I
superstring, the 10-dimensional dilaton couples differently to the
gravitational and Yang-Mills sectors because the graviton is an
excitation of closed strings while the Yang-Mills fields are
excitations of open strings. For small value of the volume of the
extra-dimensions, a T-duality makes the theory equivalent to a
10-dimensional theory with Yang-Mills fields localized on a D3
brane. When compactified on an orbifold, the gauge fields couple
to fields $M_i$ living only at these orbifold points with coupling
$c_i$ which are not universal. Typically, one gets that
$M_4^2=\hbox{e}^{-2\Phi}V_6M_I^8$ while $g_{YM}^{-2}=
\hbox{e}^{-2\Phi}V_6M_I^6 +c_iM_i$. Loop corrections have also
been studied in heterotic theory by including Kaluza-Klein
excitations~\cite{dudas}. In the limit where the volume is large
compared to the mass scale, $g_{YM}^{-2}=
\hbox{e}^{-2\Phi}V_6M_H^6 -b_a(RM_H)^2/2+\ldots$. Again, they are
not universal. It follows that the 4-dimensional effective
couplings depend on the version of the string theory, on the
compactification scheme and on the dilaton.

Interestingly, while the constraints presented above assume that
only $\alpha$ was varying, it is to be expected that if it varies
then all other constants also do. In the context of unified
theories, it is possible to derive relations between the
variations of various constants, which can be used to derive
sharper observational constraints. The dominant effects~\cite{dp}
arise from the variation of the QCD scale, $\Lambda_{\rm QCD}$ and
the weak scales $v$. In particular it was
argued~\cite{calmet,dent} that $\Delta\Lambda_{\rm
QCD}/\Lambda_{\rm QCD}\sim30\Delta\alpha/\alpha$ and $\Delta
v/v\sim 80\Delta\alpha/\alpha$.

Many phenomenological models starting from the investigation by
Bekenstein~\cite{beken1} have been developed~\cite{beken4,beken5}
as well as some braneworld models with varying constant were
constructed~\cite{palma}. Damour and Polyakov~\cite{dp} proposed
to capture the features of loop corrections by modelling them as a
genus expansion, that is as a series in the string coupling
constant. The low energy action involves various couplings of the
effective four-dimensional dilaton to the different matter fields.
It follows that generically there appear couplings to matter
fields via their Yang-Mills couplings and masses as as a potential
for the scalar field. This construction includes others such as
the Bekenstein model~\cite{beken1}. Note that these models are
required in order to compare constraints at various time.

This phenomenology is related to the one of quintessence and the
light field responsible for the time variation of the constants
may also be the cosmon~\cite{wetterich}. In that sense, the
variation of the constant may shed  some light on the physics on
dark energy. Initially models of scalar-tensor
quintessence~\cite{u99,stquint} were considered. Some were using
the Damour-Nordtvedt attraction mechanism toward general
relativity to pass the Solar System constraints. The
Damour-Polyakov model~\cite{dp} was generalized~\cite{piazza02} to
a runaway potential so that the light dilaton accounts for the
variation of the constants, the acceleration of the universe and
is at the origin of a violation of the universality of free fall.

\subsection{Two Dangers}

The construction of phenomenological models that tend to explain
the  small drift of the constant in the late universe by
introducing a slow-rolling scalar field have two dangers to avoid.

First, such a light field will obey a Klein-Gordon like equation,
$\ddot\phi+3H\dot\phi=-m^2\phi+\ldots$. In order for this field
not to oscillate but still be evolving, its mass needs to be very
small, typically $m\sim H_0\sim 10^{-33}$~eV. The question arises
of the mechanism that protects it from radiative corrections, a
problem common with most quintessence models. Various solutions
have been proposed among which the possibility for this field to
be a pseudo-Goldstone boson~\cite{carroll01} or to identify this
light field with shape modulus~\cite{peloso}.

A second danger lies in the violation of the universality of free
fall due to the composition dependence of the self energy and of
the masses. This was illustrated in Bekenstein original
construction~\cite{beken1} and was further studied with
linear~\cite{beken3} and quadratic~\cite{beken2} couplings. In the
case of a light dilaton, it was shown that if it were to remain
massless then it would induce a violation of the universality of
free fall seven order of magnitudes larger than the actual
bounds~\cite{dp}. To avoid such a catastrophe, it has either to
suddenly take a mass larger than a few meV (so that gravity will
be compatible with Einstein gravity above a millimeter) or
decouple from matter~\cite{dp}. This latter mechanism is analogous
to the original Damour-Nordtvedt attraction mechanism~\cite{dn}.
Both mechanisms have different implications concerning the
variation of the coupling constants. A fixed point~\cite{wette2}
or the recently proposed chameleon mechanism~\cite{cameleon1} also
claimed to bypass this problem. A consequence is that the
improvement of the tests of the universality of free fall by 2 or
3 orders of magnitude may give some surprises.

\subsection{Varying constants in the early universe}

Inflation universally produces classical almost scale free
Gaussian inhomogeneities of any light scalars. Assuming the
coupling constants at the time of inflation depend on some light
moduli fields, it was shown~\cite{modul} that modulated
cosmological fluctuations are produced during (p)reheating. This
idea was extended to hybrid inflation~\cite{bku} where the
bifurcation value of the inflaton is modulated by the spatial
inhomogeneities of the couplings. As a result, the symmetry
breaking after inflation occurs not simultaneously in space but
with the time laps in different Hubble patches inherited from the
long-wavelength moduli inhomogeneities. In this model, the
consistency relation of inflation is modified. These light field
can also be at the origin of non-Gaussianity~\cite{ng}.

If couplings depend on the value of some light fields, then they
have most probably developed super-Hubble correlations of typical
amplitude $10^{-5}$, simply because the quantum fluctuations of
any light field are amplified during inflation. It follows that
one expects spatial variation with these correlation (see e.g.
Ref.~\cite{piazza02}). They may have some observational
effects~\cite{mota}, in particular on the CMB~\cite{sigurdson}
polarization and Gaussianity.

\section{Conclusions}

In conclusion, testing for the variation of fundamental constants
is a test of fundamental physics, and in particular of general
relativity. It completes other tests that can be applied on
cosmological scales. Such tests are needed to substantiate the
physics of the dark sector that plays an increasing role in
cosmology.

While observational constraints become sharper, the debate on
$\alpha$ is not over yet but recent observations have not been
able to reproduce the detection of a lower $\alpha$ in the
past~\cite{webb99,webb01,murphy03}. More important, there now
exists a physical model, requiring no new physics, to interpret
the supposed variation of $\alpha$ as an enhancement of heavy
isotopes of magnesium. We can now hope that the debate will be
settled in the coming years.

The future will offer new constraints, particularly with atomic
clocks in space (ACES), new tests of the universality of free fall
(recent launch of $\mu$SCOPE) as well as new methods, as
illustrated by the recent activity. These developments will
probably shed some light on a possible scalar field acting in the
late universe or on some new structures such as higher dimensions.\\

{\bf Acknowledgments}
 I thank G. Esposito-Far\`ese, E. Flam, P. Petitjean and C. Schimd for
 many discussions on this subject,
 as well as my collaborators in some of the works presented here,
 N. Aghanim, F. Bernardeau and  Y. Mellier. I am grateful  to
 the organizers of the workshop for their kind invitation.

\end{document}